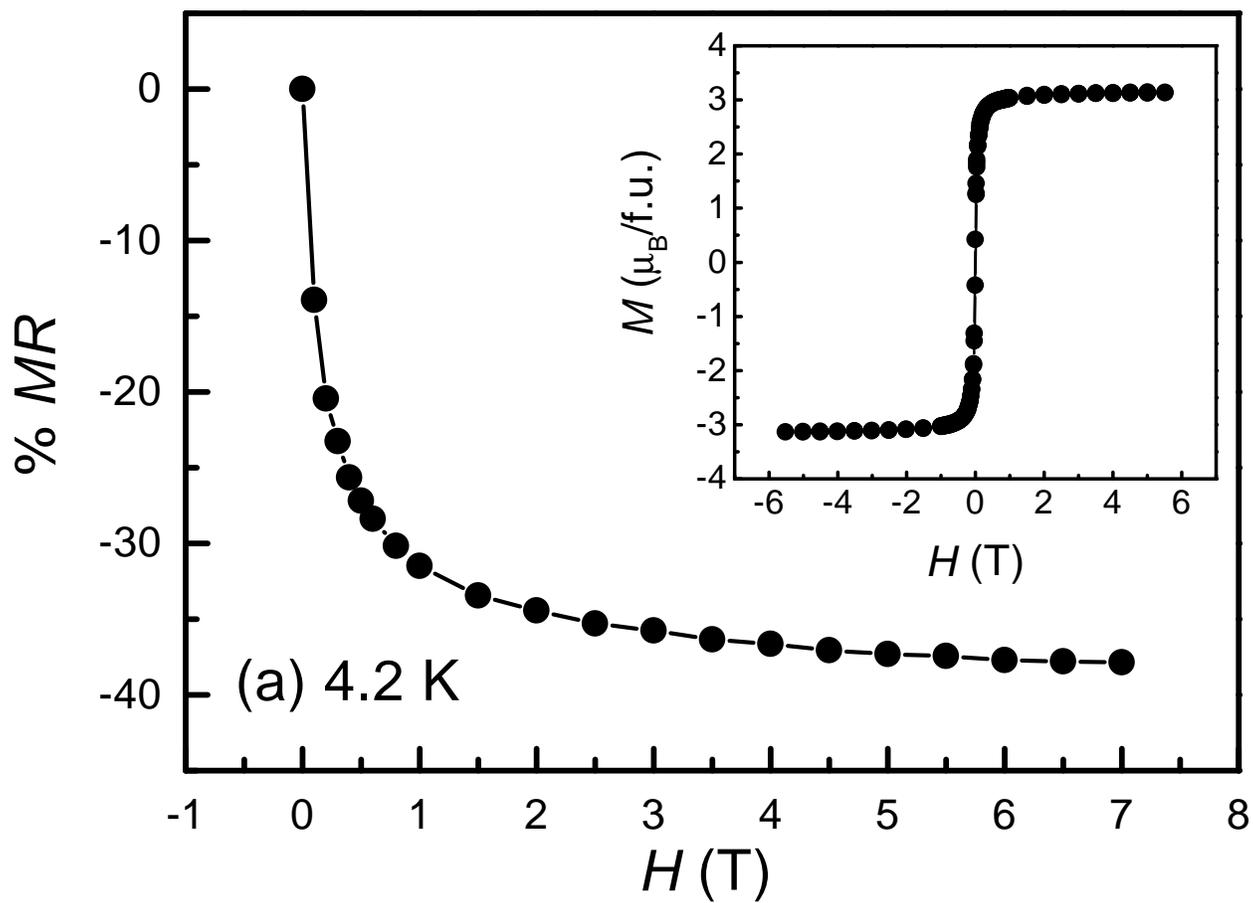
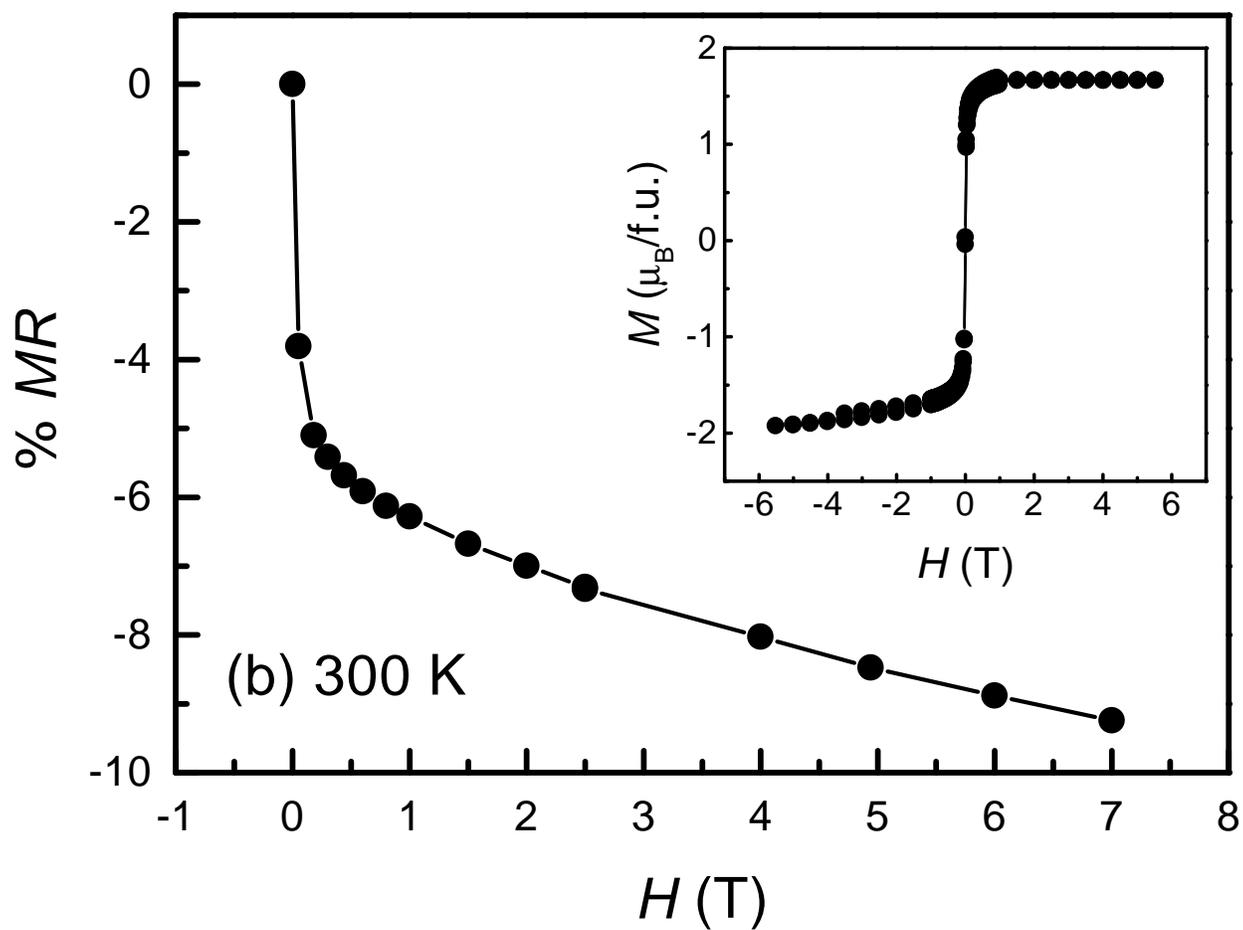

Fig. 1

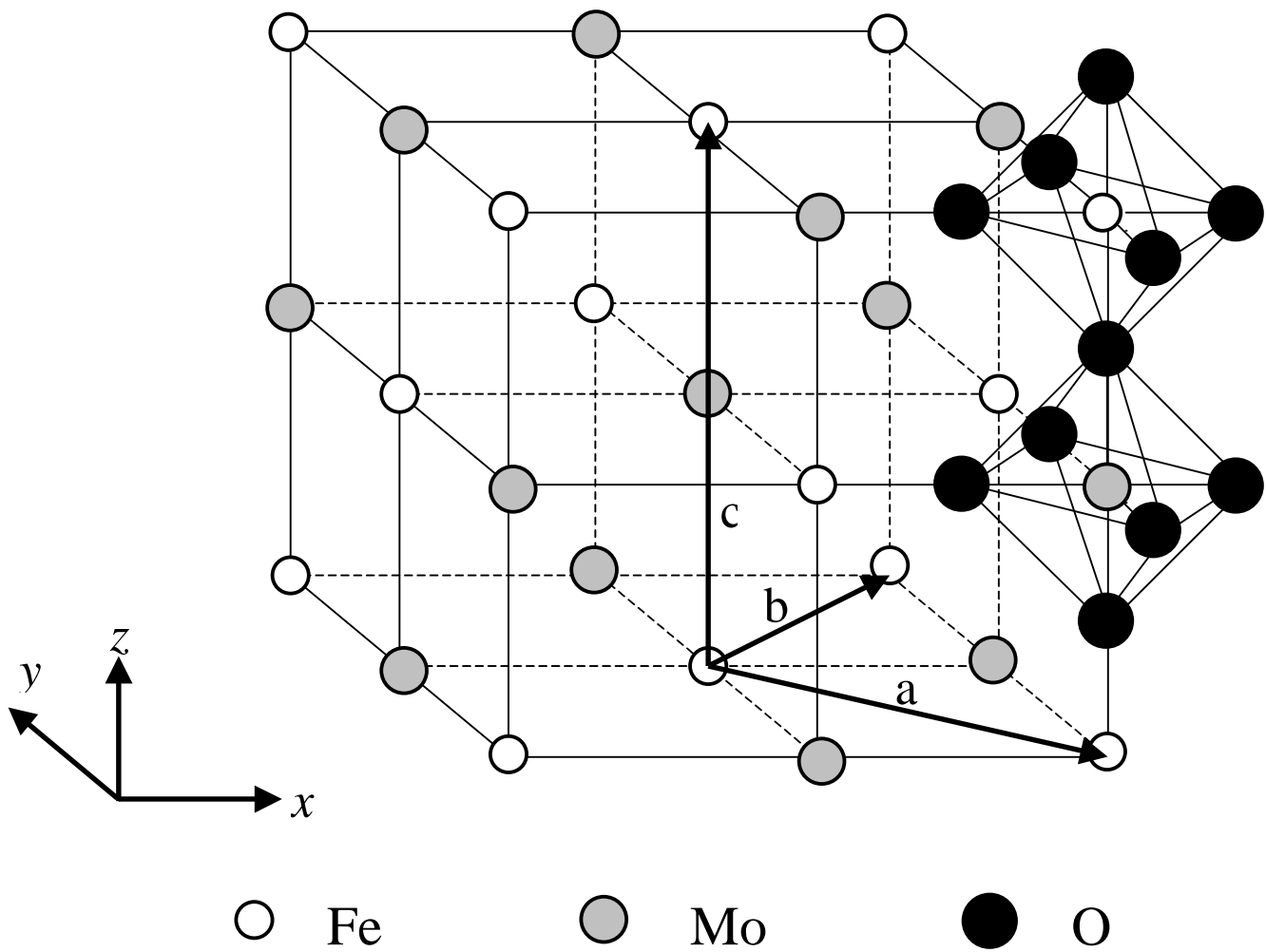

Fig. 2

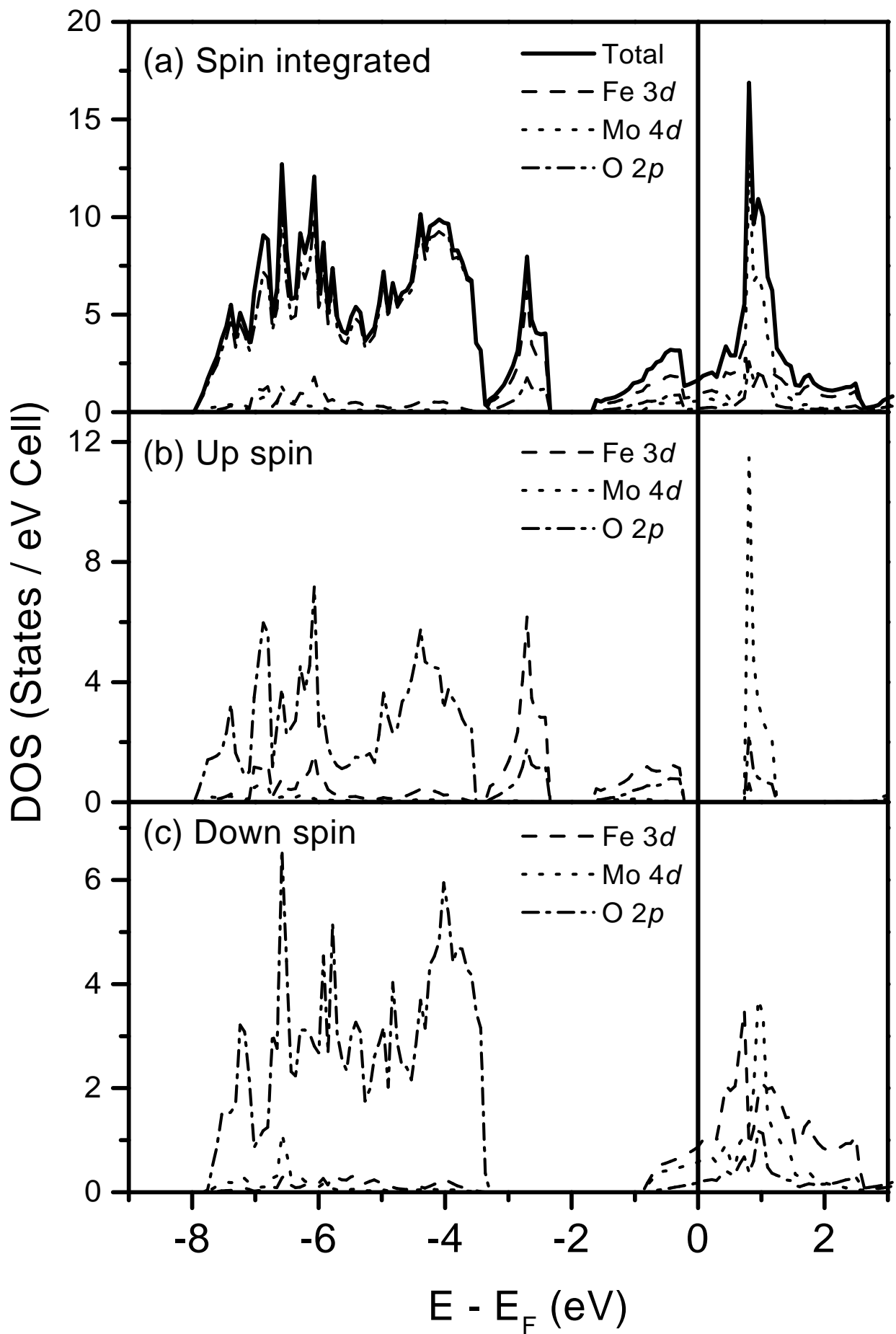

Fig. 3

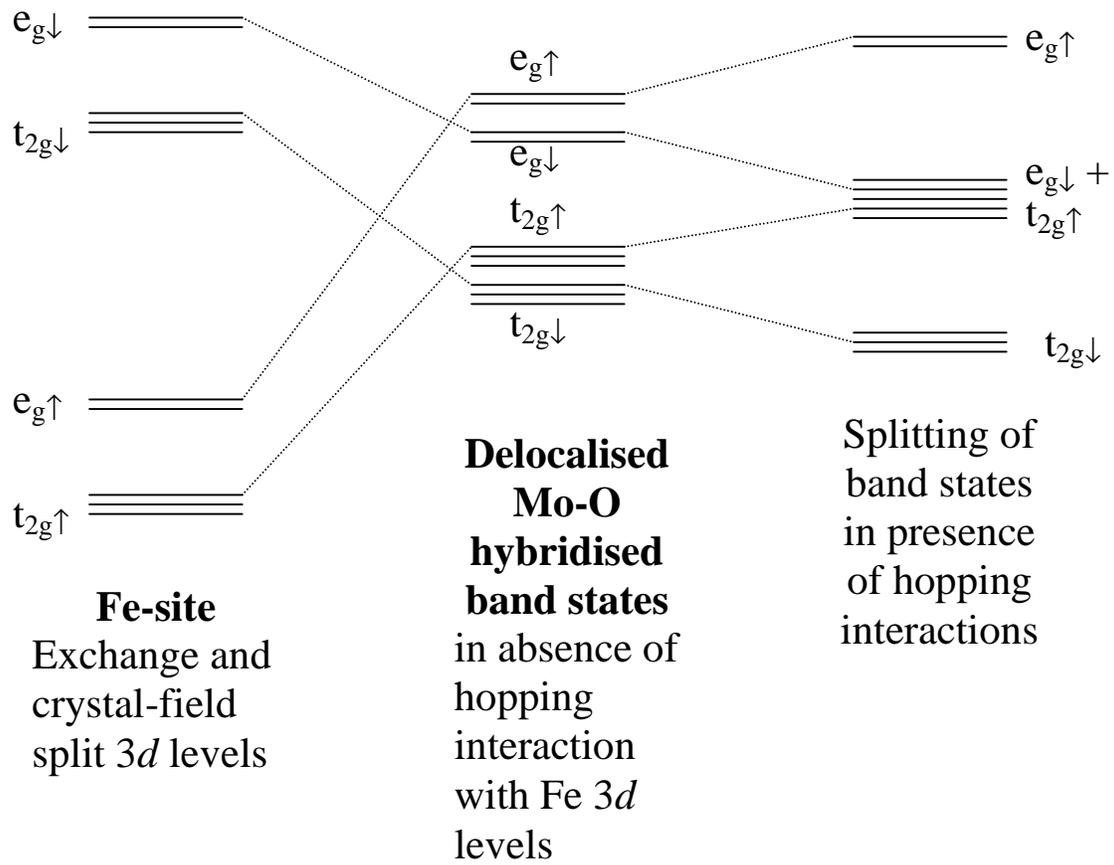

Fig. 4

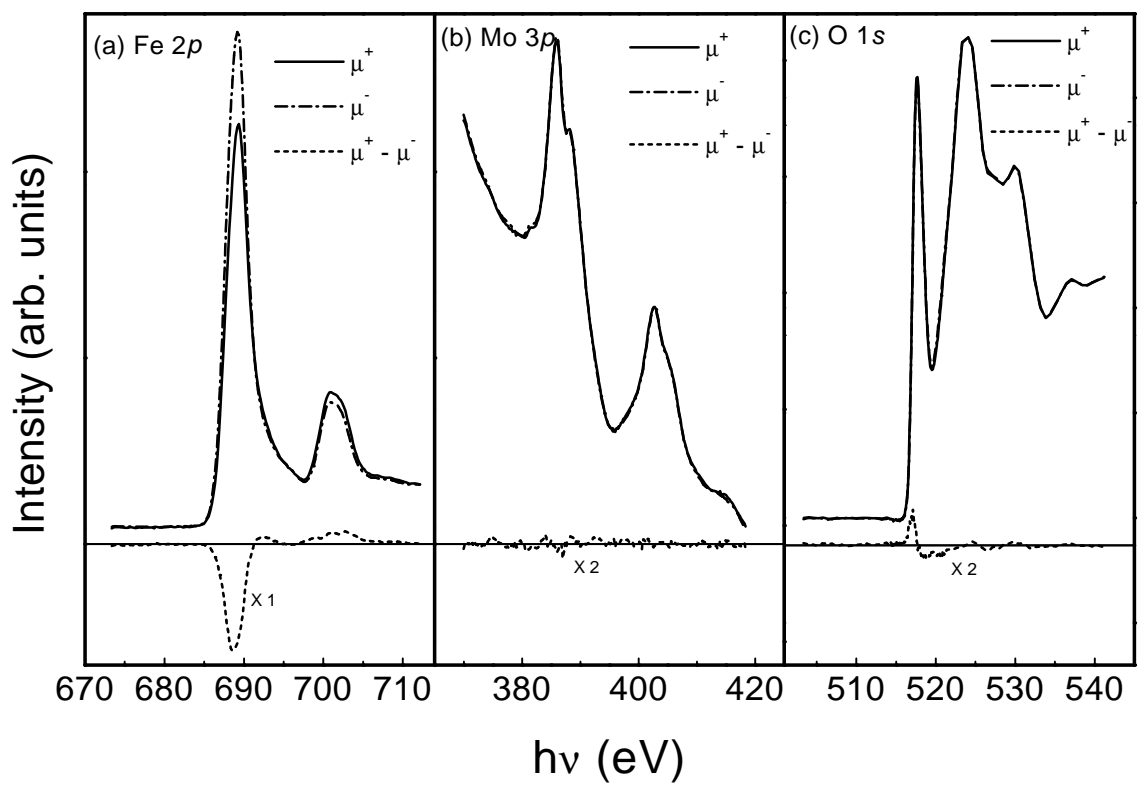

Fig. 5

**A new class of magnetic materials: $Sr_2FeMoO_6$ and related compounds**

D.D. Sarma

Solid State and Structural Chemistry Unit

Indian Institute of Science

Bangalore 560 012

INDIA

Phone: ++91-80-309 2945, Fax: ++91-80-360 1310

E-mail: sarma@sscu.iisc.ernet.in

Also at Jawaharlal Nehru Centre for Advanced Scientific Research, Bangalore.

**Abstract:**

Ordered double perovskite oxides of the general formula, $A_2BB'O_6$, have been known for several decades to have interesting electronic and magnetic properties. However, a recent report of a spectacular negative magnetoresistance effect in a specific member of this family, namely $Sr_2FeMoO_6$, has brought this class of compounds under intense scrutiny. It is now believed that the origin of magnetism in this class of compounds is based on a novel kinetically-driven mechanism. This new mechanism is also likely to be responsible for the unusually high temperature ferromagnetism in several other systems, such as dilute magnetic semiconductors, as well as in various half-metallic ferromagnetic systems, such as Heussler alloys.

**Introduction:**

In recent times, there has been a spectacular increase in research activities related to doped manganites, sparked by the observation of a remarkable decrease of resistance in such samples on the application of a magnetic field [*1]. This negative magnetoresistance, now known as the colossal magnetoresistance (CMR) because of the spectacularly large effect, is potentially useful in magnetic storage devices. Experimental and theoretical efforts have now established a strong coupling of electronic, magnetic and structural degrees of freedom as being responsible for the CMR properties in the manganites [*2-4]. Subtle interplay of theses interactions gives rise to a wide spectrum of interesting physical properties in terms of charge and orbital ordering in addition to CMR properties in doped manganites [5]. While the study of such doped manganites has been most rewarding in terms of various fundamental issues, there are two main factors that undermine its wide spread technological use. These are the low temperature and the high magnetic field usually required to have an appreciable negative magnetoresistance response from these manganites. Since the CMR effect is most significant close to the magnetic ordering temperatures, there has been an intense search for compounds with magnetic ordering temperatures substantially higher than the $T_c$ (~ 200-350 K) in manganites. Recently, it has been reported [**6] that $Sr_2FeMoO_6$, an ordered double perovskite of the general formula $A_2BB'O_6$ and containing no manganese, has a $T_c$ of about 415 K, indicating a larger interatomic exchange coupling strength, and exhibits a pronounced negative CMR at lower magnetic fields and higher temperatures compared to the doped manganites.

Besides the technologically desirable attributes of a more pronounced CMR response at higher temperatures and lower fields, there are some important fundamental aspects that distinguish $Sr_2FeMoO_6$ from the doped manganites. This system is an undoped one and its lattice does not appear to play any significant role, in contrast to the manganites. These facts would suggest that $Sr_2FeMoO_6$ is a simpler system to understand its physical properties in detailed theoretical terms. Surprisingly, in spite of this apparent simplicity, there are many open issues of fundamental importance concerning the electronic and magnetic structures of this compound. The most basic of all the unexpected properties of $Sr_2FeMoO_6$ is the occurrence of such a high magnetic transition temperature. It is unusual in view of the fact that the magnetic $Fe^{3+}$ ions are far separated in this compound, thereby suggesting a weak magnetic interaction. Moreover, such interactions between $3d^5$ ions mediated *via* other nonmagnetic ions are expected to be antiferromagnetic due to the superexchange mechanism. This expectation is supported by the observation of an antiferromagnetic ground state of the closely related system, $Sr_2FeWO_6$, with a Néel temperature, $T_N \approx 37$ K [7]. Thus, a $T_c$ of about 415 K in $Sr_2FeMoO_6$, which is higher than even that in the manganites, suggests a novel origin of magnetism in this compound. It is important to note here that there are several other examples of both ferromagnetic and antiferromagnetic compounds within the $A_2BB'O_6$ double perovskite family of compounds; for example, $Sr_2FeReO_6$ and $Sr_2CrMoO_6$ are ferromagnetic, while $Sr_2MnMoO_6$ and $Sr_2CoMoO6$ are antiferromagnetic [8,9,10]. Thus, an explanation of the magnetic structure of $Sr_2FeMoO_6$ must also be consistent with such diverse properties observed within the double perovskite oxide systems. There are several other issues concerning the electronic and magnetic structures of this compound that are still controversial and we shall discuss some of these in this article.

**Magnetoresistance and magnetisation of $Sr_2FeMoO_6$:**

Magnetoresistance of $Sr_2FeMoO_6$ has been reported by many groups [**6, 11–14]. As a typical case adopted from ref. [11], we show in Fig. 1, the percentage magnetoresistance, *MR*, defined as

$$MR(T,H) = 100* [\rho(T,H) - \rho(T,0)]/ \rho(T,0)$$

where, *ρ(T,H)* is the resistivity of the sample at a temperature, *T* and in presence of an applied magnetic field strength of *H*. Fig. 1a shows the results obtained at *T* = 4.2 K and Fig. 1b at *T* = 300 K. We also show the magnetisation of the sample at these two temperatures as a function of the applied magnetic field in the corresponding insets. The magnetisation curve, exhibiting typical hysteresis, establishes the system to be magnetic even at the room temperature. At both these temperatures, the sample is characterised by sharp and pronounced magnetoresistive responses in the low-field regime, though the magnitude of the MR is considerably higher at the lower temperature. Beyond 1 Tesla, the MR exhibits a slower change without showing any sign of saturation up to the highest magnetic field (7 Tesla). The MR changes significantly, by about 6.5% at 4.2 K and 3% at 300 K, in the larger field region between 1 and 7 Tesla. The low field response is most likely contributed by the spin scattering across different magnetic domains in these polycrystalline samples. This conclusion is supported by an absence of the sharp low-field MR response in single crystalline bulk [15] and epitaxial [16–18] samples of $Sr_2FeMoO_6$.

**Crystal and electronic structure of $Sr_2FeMoO_6$:**

The crystal structure of $Sr_2FeMoO_6$ is close to that of an ordered double perovskite structure. We show a schematic figure of this structure in Fig. 2. The unit cell dimensions are **a** = **b** = 5.57 Å, and **c** = 7.90 Å with a space group of I4/mmm [**6,11,19], indicating a small distortion from the idealised cubic structure [20]. As can be seen from this figure, Fe and Mo sites alternate at the cube corner positions, separated by intervening oxygen ions at the edge-centre positions. Extensive band structure calculations have been carried out to understand the electronic and magnetic structures of this compound [**6,**21,*22]. I show results of a typical calculation of the density of states (DOS) along with the partial Fe *d*, Mo *d* and O *p* DOS in Fig. 3. The spin integrated DOS and partial DOS are shown in Fig. 3a, while the corresponding spin-up and spin-down components are shown in Fig. 3b and 3c, respectively. It can be easily seen from Fig. 3b that there is a substantial gap in the spin-up DOS across the Fermi energy, $E_F$. In contrast to this, the spin-down channel shows finite and continuous DOS across the $E_F$ in agreement with the metallic state of this system. Thus, these results suggest that ordered $Sr_2FeMoO_6$ has a half-metallic ferromagnetic ground state, where one spin channel (the up-spin channel) behaves like an insulator with a finite gap at $E_F$, while the other spin channel (the down-spin one) has finite DOS at $E_F$. The most important consequence of this is that the mobile charge carriers in this system are fully spin-polarised. Such a complete spin-polarisation is known to be essential for the pronounced CMR effect observed in doped manganites (see Fig. 1) [23,24]. However, as we shall discuss later, the effect of disorder, inevitably present in these systems, tend to destroy the half-metallic state [*22] with important consequences on the CMR properties [11].

**Basic considerations of the magnetic structure in $Sr_2FeMoO_6$:**

Based on such band structure results, it has been suggested [**6] that this compound consists of $Fe^{3+}$ $3d^5$ $S=5/2$ and $Mo^{5+}$ $4d^1$ $S=1/2$ ions alternating along the cubic axes. The Fe and Mo sublattices are ferromagnetically coupled within each sublattice, while the two sublattices are supposed to be antiferromagnetically coupled to give rise to a $S=2$ state. Different mechanisms have been suggested for the observed magnetic structure. In close analogy to the case of manganites, it has been often suggested [13,25,26] that a double exchange mechanism [27,28] is responsible for the ferromagnetic coupling between the Fe sites. In this scenario, the delocalised electron contributed by Mo $4d^1$ configuration plays the role of the delocalised $e_g$ electron in the manganites. There are, however, some important distinctions between the physics of the manganites and that of $Sr_2FeMoO_6$. In the case of the former, both the delocalised $e_g$ electron and the localised $t_{2g}$ electrons reside at the *same* site, namely the Mn sites. The spin moment of $t_{2g}^3$ localised states couples ferromagnetically to the spin of the $e_g^1$ delocalised electron, due to the *intra-atomic Hund's coupling strength, I,* arising from the exchange stabilisation of the parallel spin arrangement. In the case of $Sr_2FeMoO_6$, while the delocalised electron at the Mo site and the localised electrons at the Fe sites are nominally at two different sites, band structure results in Fig. 3c suggest that the mobile electrons also have a finite Fe character due to sizable hopping interaction strengths coupling Fe $d$ states to Mo $d$ states *via* the oxygen $p$ orbitals. This might appear to support a double exchange mechanism. However, the localised up-spin orbitals at the Fe site in $Sr_2FeMoO_6$ are already fully-filled, making it impossible for another up-spin electron to hop to the Fe site and forcing the delocalised electron to be down-spin. Therefore, Hund's coupling strength, $I$, between the parallely oriented localised and delocalised electrons, which provides the energy scale of the on-site spin coupling in the double exchange mechanism for the manganites, is irrelevant and cannot be invoked in the case of these double perovskites. This shows that the *antiferromagnetic* coupling between the localised and the delocalised electrons, which must be substantial to yield such a large $T_c$ in $Sr_2FeMoO_6$, originates from a totally different mechanism.

In a recent work [29], ferromagnetic $T_c$ has been calculated within a double exchange-type Hamiltonian for $Sr_2FeMoO_6$, but assuming an antiferromagnetic coupling between the localised and delocalised spins. It is to be noticed here that either a ferromagnetic or an antiferromagnetic coupling (or any other well-defined coupling) between the delocalised electrons in the system and the localised electrons at each Fe site will lead to a ferromagnetic ordering of the Fe sublattice. Thus, understanding the ferromagnetic ordering of Fe ions in $Sr_2FeMoO_6$ is nothing but understanding the nature and origin of the coupling of the mobile electrons and the localised ones in this compound. Since intra-atomic Hund's interaction strength cannot provide this coupling as already pointed out, we have to look for another mechanism to explain the occurrence of such a high $T_c$ in this compound. In some reports, it has been implicitly assumed [30] that an antiferromagnetic coupling between the Fe and the Mo sites *via* superexchange is responsible for the observed magnetic structure. This does not appear to be a very plausible scenario, since a superexchange mechanism coupling the Fe site to the delocalised and highly degenerate (five-fold degeneracy ignoring crystal-field effects) Mo *d* states will, at best, be very weak, therefore, not compatible with the unusually high ordering temperature. Moreover, it should be noted that such a superexchange mechanism requires a perfectly ordered double perovskite structure, ensuring Fe–O–Mo–O–Fe 180° interactions to give rise to a ferromagnetic coupling of the Fe sublattice. It would also suggest that Fe–O–Fe bonds, if present, will be antiferromagnetically coupled. However, recent band structure calculations using supercells to simulate mis-site disorder between Fe and Mo sites [*22] clearly show that Fe and Mo sites are invariably coupled antiferromagnetically, driving a ferromagnetic order in the Fe even for Fe–O–Fe bonds in this system. This observation effectively eliminates the possibility of superexchange interaction being the driving force for the magnetic ordering in this compound; I shall discuss the results pertaining to the disorder effect in some more detail in a later section. These considerations also suggest that this compound should not be considered as a ferrimagnet, but a ferromagnet.

**A new mechanism of magnetic interactions and the origin of magnetism in $Sr_2FeMoO_6$ and related compounds:**

A well-defined spin ordering between the delocalised electrons and the localised Fe electrons presupposes a large spin splitting of the delocalised band, derived from the Mo $d$ and oxygen $p$ states. This is surprising in view of the fact that Mo is usually not a strongly correlated system and, consequently, a magnetic moment at the Mo site is a rarity. However, detailed band structure calculations [**21] show that the nominally Mo $d$ band in this compound exhibits an exchange splitting that is larger than the bandwidth. A novel mechanism has been recently proposed to explain this new type of magnetic interaction between the localised electrons and the conduction electrons, leading to a strong polarisation of the mobile charge carriers [**21]. Following the arguments presented in this work, we explain the origin of this large spin splitting of the effective Mo $d$ band with the help of the schematic shown in Fig. 4, where we plot the relevant energy level scheme for this compound. $Fe^{3+}$ $3d^5$ configuration is known to have a large exchange splitting of the $d$ level in spin-up $3d•$ and spin-down $3d•$ states [**21,31]. There is a further crystal field splitting of the $3d$ states in terms of $t_{2g}$ and $e_g$ states in the octahedral symmetry of the Fe ions [32], though the crystal-field splitting in the case of $Fe^{3+}$ is considerably smaller than the exchange splitting. This is shown on the left-hand side of the schematic in Fig. 4. Nonmagnetic band structure results suggest that the Mo $4d$–O $2p$ hybridised states appear at about 1.4 eV above the Fe $d$ states. The exchange splitting of these states is expected to be very small, though it has a substantial crystal-field splitting. This is shown in the middle panel of Fig. 4. This would be the scenario in absence of any hopping interactions coupling the Fe states to the delocalised states derived from the Mo 4$d$–O 2$p$ states. In the presence of hopping interactions, there is a finite coupling between the states of the same symmetry and spin at the Fe and the delocalised electrons. This hopping interaction leads to a substantial admixture of Fe $d$ contribution in the nominally Mo 4$d$–O 2$p$ derived delocalised states as seen in Fig. 3c. But more importantly, it leads to shifts in the bare energy levels. It is then easily seen that the delocalised $t_{2g}$● states will be pushed up and the $t_{2g}$● states will be pushed farther down by hybridisation with the corresponding Fe states, as shown in the figure. There will be shifts in the delocalised $e_g$ levels also, though it is not relevant for the mechanism we are discussing here. The opposite shifts of the up- and down-spin conduction states, therefore, induce a spin-polarisation of the mobile electrons due to purely hopping interactions between the localised electrons and the conduction states. This kinetic energy driven mechanism [**21] obviously leads to an antiferromagnetic coupling between the localised and the conduction electrons, since the energy is lowered by populating the down-spin conduction band with respect to the majority spin orientation of the localised electrons. The extent of the spin-polarisation of the conduction electrons derived from this mechanism [**21] is primarily governed by the effective hopping strength and the charge-transfer energy between the localised and the delocalised states, as also has been suggested in a subsequent paper [33] based on perturbative arguments. Moreover, the effective antiferromagnetic coupling strength between the spins of the localised and the delocalised electrons is also dependent on these two parameters. Detailed many-body calculations [**21] have shown that the spin-polarisation of the conduction band in Sr$_2$FeMoO$_6$ is as large as 1—1.5 eV and the strength of the antiferromagnetic coupling between the conduction band and the localised electrons at the Fe site is of the order of 18 meV which is larger than that in the doped manganites, explaining the high $T_c$ in this compound.

It is clear that this mechanism will be operative whenever the conduction band is placed within the energy gap formed by the large exchange splitting of the localised electrons at the transition metal site. However, if the band generated from Mo 4$d$–O 2$p$ states were to be outside this gap, both up- and down- states will be shifted in the same sense and the large energy gain *via* the antiferromagnetic coupling will not be possible. This is believed to be the case in $Sr_2FeWO_6$, where the strong hybridisation between the W 5$d$ and the O 2$p$ states drives the hybridised states above the $t_{2g}\bullet$ level of Fe [34]. Such an energy level scheme then cannot stabilise the antiferromagnetic coupling between the electron in the delocalised states and the localised ones; instead it transfers the electron from the W 5$d$–O 2$p$ hybridised state to the Fe 3$d$ level, leading to an insulating compound with formally $W^{6+}$ and $Fe^{2+}$ states. In absence of any mobile electrons, the $Fe^{2+}$ sites couple *via* superexchange to give rise to an antiferromagnetic insulating state in $Sr_2FeWO_6$, in contrast to the metallic ferromagnetic state of $Sr_2FeMoO_6$. This new mechanism of magnetism proposed in ref. [**21] can also explain the metallic ferromagnetic ground state of $Sr_2FeReO_6$ as well as the antiferromagnetic state of $Sr_2CoMoO_6$ and $Sr_2MnMoO_6$ [33,34]. It has also been suggested [33] that this mechanism [**21] is possibly responsible for the magnetism in a host of other compounds, such as $In_{1-x}Mn_xAs$, $V(TCNE)_2.1/2CH_2Cl_2$. Dilute magnetic semiconductors, such as $Ga_{1-x}Mn_xAS$ are also examples of this new class of magnetic systems. In all these systems, the conduction band is polarised antiferromagnetically with respect to the localised moment at the transition metal site due to the hopping interactions between the two and a large exchange splitting of the localised state, thereby driving a ferromagnetic arrangement of the localised moments. It is also likely that this type of magnetic interactions are operational in other half metallic ferromagnetic systems, such as Heussler alloys. Having discussed the nature and origin of the magnetic coupling in these systems, we now turn to some specific issues concerning the electronic and magnetic structures of $Sr_2FeMoO_6$ system, which has generated some debate and controversies in the recent literature [**6,11,26,35-40].

**Some details of the magnetic and electronic structures of $Sr_2FeMoO_6$:**
The original suggestion [**6] of the magnetic structure in terms of a ferrimagnetic arrangement of $Fe^{3+}$ $S=5/2$ and $Mo^{5+}$ $S=1/2$ states was questioned in an early work [26] on the basis of neutron measurements, since no measurable moment could be observed at the Mo sites. A subsequent investigation, however, claimed to find the expected moment of 1 $\mu_B$ at the Mo site [36]. Such conflicting conclusions based on the same technique suggest that either the neutron data is somewhat insensitive to the small moment at the Mo site in this case or its interpretation is model dependent, leading to different claims by different groups. In order to probe the magnetic moments at the Mo as well as the Fe and the O sites directly, X-ray magnetic circular dichroism (XMCD) has been employed recently [37]. It is well established that this technique has the ability to provide information concerning the site and angular momentum specific contributions to the magnetic moment [41-43]. In Fig. 5, we show the XMCD spectra at the Fe $2p \rightarrow 3d$, Mo $3p \rightarrow 4d$, and O $1s \rightarrow 2p$ edges from ref. [37]. While the detailed interpretations of these spectra can be found in the original paper, it is quite clear that there is a substantial dichroism in the spectra at the Fe $2p$ edge, as shown by the large intensity of the difference or the XMCD spectrum shown in the figure; this directly establishes the presence of a large magnetic moment at the Fe sites in this compound. In contrast, the dichroic signal at the Mo $3p$ edge is negligible, establishing that the magnetic moment at the Mo sites is below the detection limit ($\leq 0.2$ $\mu_B$), in agreement with the conclusions based on the neutron experiments in ref. [26], but in contrast to ref. [36]. Interestingly, there is a significant XMCD signal at the O 1$s$ edge, suggesting the presence of spin density at the O sites, instead of the Mo sites. While this is different from the originally suggested magnetic structure, the XMCD results can be easily understood in terms of the mechanism of magnetism discussed here in terms of Fig. 4. It is to be noticed here that the delocalised states are not solely Mo 4$d$ derived states, but these inevitably have a substantial admixture of O 2$p$ states. Therefore, we have referred to these states as Mo 4$d$–O 2$p$ hybrid states in our discussions. The XMCD results show that the single delocalised electron of opposite spin with respect to the Fe majority spin direction is not localised on any of the neighbouring Mo or O ions, with the spin density of this electron being spread over several sites, with a larger contribution on the six oxygens around the Fe sites and (and also possibly in the down-spin channel of the Fe site), with considerably smaller spin density at the Mo sites. Thus, it appears that the delocalised spin density, antiferromagnetically coupled to the localised spins at the Fe sites, prefers to be spatially closer to the central Fe sites, thereby gaining a stronger antiferromagnetic coupling between the localised and the delocalised spins rather than residing at the farther Mo sites.

The other controversial issue concerning the electronic structure of this compound has been the determination of the formal valence state of Fe, which in turn determines the valence state of Mo *via* charge neutrality. Originally, it was suggested that Fe is in the 3+ state [**6,11,37]. However, several reports have appeared [40,44] claiming that the state of Fe is closer to 2.5+. Some of these are based on the analysis of Mössbauer data, suggesting that the minority $d$ band at the Fe site is also occupied. This can also be seen easily in all band structure results (see Fig. 3), clearly showing the contribution of Fe $d$• states in the occupied part near the Fermi energy. However, this does not imply that Fe is in a fractional valent state, as has often been suggested in the past. The occupancy of the minority $d$ states occurs *via* covalent mixing of these states with other states, such as the O $2p$ and Mo $4d$ states, due to the presence of a large hopping interaction strength (see schematic in Fig. 4). In such covalent systems in presence of strong correlation effects, the only way one can associate a formal valency with a specific site is by an analysis of the ground state. While it is impossible to describe the full many-body ground state of such a system, one often attempts to describe the observed spectroscopic properties in terms of a representative fragment of the solid; this approach is known as the cluster approximation [45-47], which has been very useful in describing a large number of transition metal compounds [48]. The typical fragment in the case of $Sr_2FeMoO_6$ pertaining to the Fe site is the $FeO_6$ octahedron. If Fe is in $Fe^{n+}$ valence state with each oxygen in $O^{2-}$ state, this cluster can be written as $(FeO_6)^{n-12}$ cluster, ($n$-12) being the uncompensated charge on this cluster. The extreme ionic configuration of this cluster corresponds to an electron configuration state of $|3d^{8-n} 2p^{36}>$ with (8-$n$) electrons in the Fe $3d$ and 36 electrons in the $2p$ orbitals of six oxygen sites. In presence of finite hopping interactions, the ground state wave-function will be a linear combination of all possible states generated from this ionic state by hopping of electrons from the O $2p$ to Fe $3d$ states, such as $|3d^{9-n} 2p^{35}>$, $|3d^{10-n} 2p^{34}>$, *etc*. It is then obvious that the expectation value of the $d$-occupancy, $<n_d>$, will necessarily be larger than (8-$n$) due to the covalent mixing of the higher lying states connected *via* hopping; however, this does not imply that the valency of the Fe site in this case is less than $n$+. In this sense, the question of formal valency is really related to the question of which ionic configuration constitutes a better approximation as *one of the basis states* to expand the ground state wave-function. In the specific context of $Sr_2FeMoO_6$, this reduces to the question whether the typical fragment $FeO_6$ is better represented by linear combinations of $|3d^5 2p^{36}>$ combined with its excited states

(e.g. $|3d^6\ 2p^{35}\rangle$, $|3d^7\ 2p^{34}\rangle$) suggesting a $Fe^{3+}$ valency, or by linear combinations of $|3d^6\ 2p^{36}\rangle$ combined with its excited states (e.g. $|3d^7\ 2p^{35}\rangle$, $|3d^8\ 2p^{34}\rangle$, etc.) indicating a formal $Fe^{2+}$ valence state. A recent x-ray absorption study at the Fe 2p → 3d threshold [37] clearly shows that the spectrum is only compatible with the former ($Fe^{3+}$) configuration and is incompatible with the latter ($Fe^{2+}$) configuration. However, it is important to note here that such a description is only an approximate one, since the cluster considered here will couple to the rest of the solid via hopping interactions and therefore its total electron occupancy cannot be assumed to be an integral one. This problem, though less severe in the case of highly insulating systems, becomes quite important for systems with delocalised electrons as in the case of $Sr_2FeMoO_6$.

**Effect of mis-site disorder on the magnetic properties and its implications:**

Since the ionic sizes of $Fe^{3+}$ and $Mo^{5+}$ are similar, there is a finite concentration of mis-site disorder in $Sr_2FeMoO_6$, which interchanges the positions of Fe and Mo sites in a random fashion. The first preparation of extensively disordered $Sr_2FeMoO_6$ with an ordering of about 30% was achieved by melt-quenching the sample [11]. Subsequently, there have been publications [44] reporting the synthesis of samples with different degrees of ordering of the Fe and Mo sites. The most significant effect of disordering in this system is to reduce the net magnetisation of the sample. This observation is relevant in the context of the normally prepared so-called ordered sample, which usually has an ordering of about 90% [**6,11]. Such samples invariably exhibit a saturation magnetisation in the order of 3.1 $\mu_B$ per formula unit (f.u.) where as the expected value considering the electron count and the magnetic structure is 4 $\mu_B$/f.u.; the decrease of the observed magnetisation with respect to the expected one has been ascribed to the finite concentration of mis-site disorder [49,50]. However, the nature and origin of this decrease of the magnetisation in presence of disorder is still a matter of debate in the literature [18,44]. There are two distinct ways that the net magnetisation may be reduced in $Sr_2FeMoO_6$ in presence of mis-site disorders. One possibility is that the disorder destroys the specific spin arrangement of Fe and Mo sublattices without any significant effect on the individual magnetic moments at these sites. This can be achieved by transforming the ferromagnetic coupling between some of the Fe sites to an antiferromagnetic coupling. This view has been preferred by most in recent time, under the assumption that Fe-O-Fe interactions, induced by the mis-site disorder in place of Fe-O-Mo, will be antiferromagnetic driven by the superexchange. Alternately, the magnetic moments at each individual site may decrease due to the different chemical environment induced by the disorder, without affecting the nature of the spin order within the Fe and Mo sublattices. The real situation may even be a combination of both these effects, with a simultaneous reduction in the magnetic moments at different sites as well as a change in the nature of the magnetic coupling between different sites. Recently, extensive *ab initio* band structure calculations [*22] with supercells to simulate mis-site disorders between Fe and Mo have clearly shown that the Fe sites continue to be ferromagnetically coupled in every case including where the bonding contains Fe-O-Fe units. It has also shown that delocalised electron generated from the Mo and O states with some admixture of Fe states also, invariably remains to be antiferromagnetically coupled to the localised Fe moments, in close analogy to the magnetic structure of the ordered system. This clearly shows that the magnetic interaction proposed in ref. [**21] and shown in the schematic Fig. 4 always dominates over the superexchange interactions in these systems. It is evident that the new mechanism discussed here, does not depend on any specific geometry or the lattice of the transition metal sites, being driven by the local hopping interactions connecting Fe *d* states to the O *p* and Mo *d* states. Therefore, this interaction survives even in a disordered system, as shown explicitly by the supercell calculations [*22]. This insensitivity to the specific atomic arrangement of this mechanism also makes it the most plausible candidate to explain magnetism in dilute magnetic semiconductors, where the magnetic ions are substituted randomly in the lattice of the semiconductor. These supercell calculations establish that the decrease in the magnetic moment in presence of disorder arises solely from a decrease of the individual moments at the Fe sites due to the change in the chemical environment and can be understood in terms of the local electronic structure around each of the inequivalent Fe sites [*22], in contrast to the prevalent view in the literature at this time.

**Conclusions:**

I have discussed the interesting physical properties of the double perovskite, $Sr_2FeMoO_6$, mainly in terms of magnetisation and magnetoresistance behaviours. After pointing out that the usual magnetic interactions operative in most of the transition metal compounds cannot account for the observed magnetism in this compound, I have described a new magnetic interaction responsible for the unusual magnetic structure in this compound. It turns out that this novel mechanism is also responsible for magnetism in a large number of seemingly unrelated systems, such as dilute magnetic semiconductors and Heussler alloys.

**Acknowledgement:** Besides the published (and also unpublished) literature, I have gained immensely from extensive discussions with my colleagues: T.N. Guru Row, H.R. Krishnamurthy, A. Kumar, P. Mahadevan, S. Majumdar, R. Nagarajan, G. Nalini, S. Ray, T. Saha-Dasgupta, E.V. Sampathkumaran and B. Sriram Shastry. I thankfully acknowledge the help of S. Ray and T. Saha-Dasgupta in preparing this manuscript.

**Figure Captions:**

Fig. 1 Magnetoresistance of $Sr_2FeMoO_6$ at (a) 4.2 K and (b) 300 K as a function of the applied magnetic field. The insets show the magnetisation at these two temperatures. (Adopted from Refs. [9] and [46]).

Fig. 2 Schematic structure of $Sr_2FeMoO_6$. Only few of the oxygen atoms are shown for clarity, while the Sr atoms at the body-centre positions are not shown. The cubic axes (x,y,z) as well as the crystallographic axes (a,b,c) are also shown in the figure.

Fig. 3 Density of states (DOS) along with partial Fe $d$, Mo $d$ and O $p$ density of states are shown in three panels. The panel (a) shows the spin integrated densities, while panels (b) and (c) show the corresponding quantities for the up- and down-spin channels, respectively.

Fig. 4 Schematic of various energy level diagrams to explain the origin of the proposed magnetic interaction in $Sr_2FeMoO_6$ and related compounds. (Adopted from ref. [19]).

Fig. 5 X-ray absorption spectra at (a) Fe 2$p$, (b) Mo 3$p$ and (c) O 1$s$ edges and the corresponding x-ray magnetic circular dichroic spectra (multiplied by 2). (Adopted from ref. [34]).